\documentclass[aps,prd,preprint,superscriptaddress,tightenlines,%
nofootinbib]{revtex4}
\newcommand{\PRE}[1]{{#1}}   
\usepackage{amsmath,amssymb}
\usepackage{bm}
\usepackage{color}
\usepackage{mathrsfs}
\usepackage{epsfig}
\usepackage{latexsym}

\newcommand{\comment}[1]{}

\def\be{\begin{equation}}
\def\ee{\end{equation}}
\def\bi{\begin{itemize}}
\def\ei{\end{itemize}}
\def\ed{\end{document}}
\def\bfl{\begin{flushleft}}
\def\efl{\end{flushleft}}
\def\bea{\begin{eqnarray}}
\def\eea{\end{eqnarray}}

\def\wf{\widetilde{F}^{(0,3)}}
\def\WW{\sqrt{\left. \sigma v \right|_{WW}/3}}
\def\WW2{\left. \sigma v \right|_{WW}}

\def\coseno{\cos^2 \theta_W}
\def\s2{\sin^2 \theta_W}

\begin{document}

\preprint{
\hfil
\begin{minipage}[t]{3in}
\begin{flushright}
CERN-PH-TH/2009-220\\
FERMILAB-PUB-09-579-A
\vspace*{.4in}
\end{flushright}
\end{minipage}
}

\title{Neutralino dark matter annihilation to monoenergetic \\ gamma rays as a signal of low mass superstrings
\PRE{\vspace*{0.3in}}
}

\author{Luis A. Anchordoqui}
\affiliation{Department of Physics,\\
University of Wisconsin-Milwaukee,
 Milwaukee, WI 53201, USA
\PRE{\vspace*{.1in}}
}

\author{Haim Goldberg}
\affiliation{Department of Physics,\\
Northeastern University, Boston, MA 02115, USA
\PRE{\vspace*{.1in}}
}

\author{Dan \nolinebreak Hooper}
\affiliation{Center for Particle Astrophysics,\\ Fermi National Accelerator Laboratory, Batavia, IL 06510-0500, USA
\PRE{\vspace*{.1in}}
}

\affiliation{Department of Astronomy and Astrophysics,\\ University of Chicago, Chicago, IL 60637, USA
\PRE{\vspace*{.1in}}
}

\author{Danny Marfatia}
\affiliation{Department of Physics and Astronomy,\\
University of Kansas, Lawrence, KS 66045, USA
\PRE{\vspace*{.1in}}
}

\author{Tomasz R. Taylor}
\affiliation{Department of Physics,\\
Northeastern University, Boston, MA 02115, USA
\PRE{\vspace*{.1in}}
}

\affiliation{Department of Physics,\\ CERN Theory Division CH-1211
Geneva 23, Switzerland
\PRE{\vspace{.1in}}
}


\begin{abstract}\vskip 3mm

  \noindent We consider extensions of the standard model based on open
  strings ending on D-branes, in which gauge bosons and their
  associated gauginos exist as strings attached to stacks of D-branes,
  and chiral matter exists as strings stretching between intersecting
  D-branes. Under the assumptions that the fundamental string scale is
  in the TeV range and the theory is weakly coupled, we study models
  of supersymmetry for which signals of annihilating neutralino dark
  matter are observable. In particular, we construct a model with a
  supersymmetric R-symmetry violating (but R-parity conserving)
  effective Lagrangian that allows for the $s$-wave annihilation of
  neutralinos, once gauginos acquire mass through an unspecified
  mechanism. The model yields bino-like neutralinos (with the measured
  relic abundance) that annihilate to a $\gamma \gamma$ final state
  with a substantial branching fraction ($\sim$ 10\%) that is orders
  of magnitude larger than in the minimal supersymmetric standard
  model.  A very bright gamma-ray spectral line could be observed by
  gamma-ray telescopes.

\end{abstract}

\maketitle

Superstring theory is a promising candidate to explain the
underlying symmetries of nature, e.g., the probable existence and
breaking of supersymmetry (SUSY). In particular, TeV-scale superstring
theory provides a brane-world description of the standard model, which
is localized on hyperplanes extending in $p+3$ spatial dimensions, the
so-called D-branes. Gauge interactions emerge as excitations of open
strings with endpoints attached on the D-branes, whereas gravitational
interactions are described by closed strings that can propagate in all
nine spatial dimensions of string theory (these comprise parallel
dimensions extended along the $(p+3)$-branes and transverse
dimensions).  The apparent weakness of gravity at energies below a few
TeV can then be understood as a consequence of the gravitational
force ``leaking'' into the transverse compact dimensions of
spacetime. This is possible only if the intrinsic scale of string
excitations is also of order a few TeV. Should nature be so
cooperative, one would expect to see a few string states produced at
the LHC, most distinctly manifest in the dijet~\cite{Meade:2007sz} and
$\gamma + $jet~\cite{Anchordoqui:2007da} spectra resulting from their
decay.

An attractive feature of broken SUSY is that with R-parity
conservation the lightest supersymmetric particle (LSP) is a possible
candidate for cold dark matter~\cite{Goldberg:1983nd}.  Requiring the
relic abundance to conform with cosmological dark matter measurements
serves to constrain the underlying theory. A consequence is that
it may be possible to detect the annihilation products of such particles,
such as gamma rays, charged leptons, and neutrinos.

In this Letter, we propose new processes, based in brane-world string
theory, for the efficient annihilation of neutralino LSP's
($\chi^0$'s) into monochromatic gamma rays, $Z$-bosons, charged
$W$s and pairs of gluons (via $\chi^0 \chi^0 \to \gamma \gamma,\, \gamma Z,\, ZZ,\,
W^+W^-,\,{\rm and}\ gg$). By requiring that the total annihilation rate generate
the measured dark matter abundance~\cite{Komatsu:2008hk}, we constrain the parameters of the
model: the string scale, the neutralino mass, the string coupling
constant, and two unknown dimensionless parameters that depend
on the details of compactification. We then calculate the gamma-ray 
spectrum from neutralino annihilation in the central
region of the Milky Way, and explore the prospects for discovery
with present and
future gamma-ray telescopes.

The basic unit of gauge invariance for D-brane constructions is a
$U(1)$ field, so that a stack of $N$ identical D-branes 
generate a $U(N)$ theory with the associated $U(N)$ gauge group.
Gauge bosons and associated gauginos (in a supersymmetric theory)
arise from strings terminating on {\em one} stack. For simplicity we
consider a model with 3 stacks corresponding to gauge groups
$U(3) \times U(2) \times U(1)$, labeled stacks $a$, $b$, and $c$,
respectively.

We consider the introduction of new operators, based on superstring
theory, that avoid $p$-wave suppression by permitting $s$-wave annihilation
into gauge bosons at an adequate rate. To create an $s$-wave, both gauginos
must be in the same helicity state, either left- or right-handed. Such
gaugino pair annihilation violates R-symmetry by $\Delta r=\pm 2$,
and is therefore forbidden in supersymmetric Yang-Mills theory, at
least at the perturbative level. However, it can appear in conjunction
with a SUSY-breaking gaugino mass generation mechanism.

\begin{figure}[tbp]
\begin{center}
\mbox{\includegraphics[height=2.3cm]{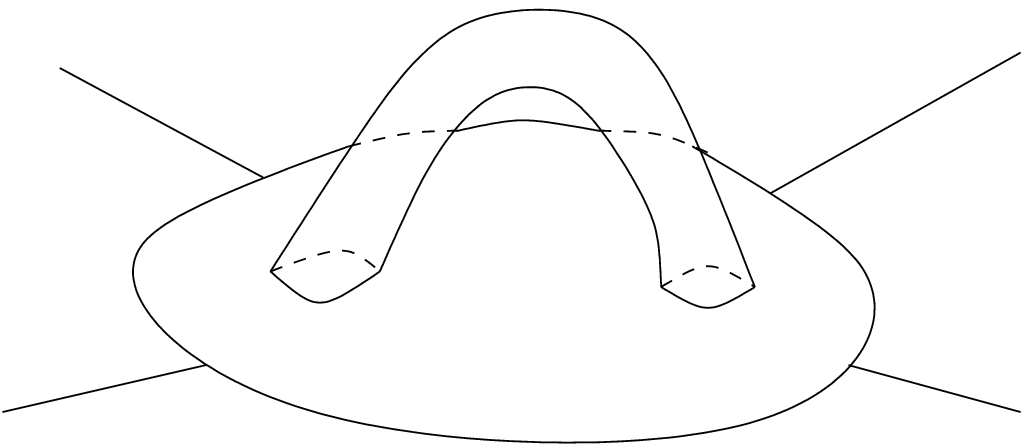}\ \ \ \ \ \ \ \ \ \ \ \ \ \ \ \ \ 
      \includegraphics[height=2.5cm]{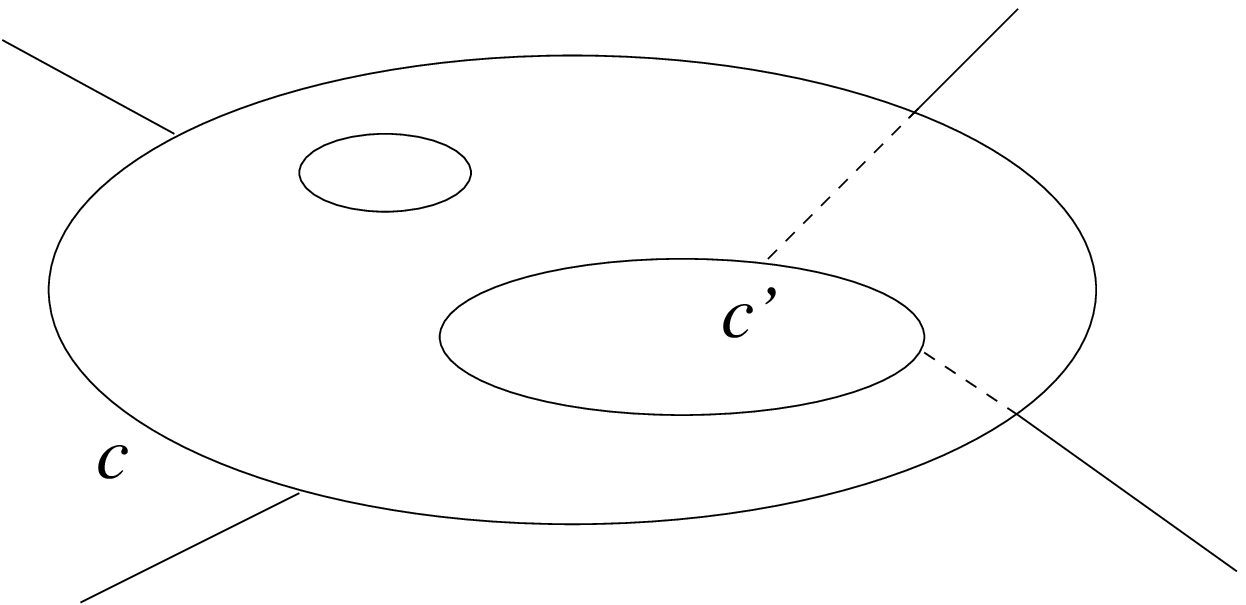}}
\caption{Left: ``Genus 3/2'' worldsheet with one handle, with four vertices inserted at the boundary.
         Right: Two-loop open string worldsheet with two vertices inserted
at the boundary $c$ and two at $c'$ while the third one is ``empty''.}
\label{fig1}
\end{center}
\end{figure}

In superstring theory, there are no conserved charges associated with
continuous global symmetries; consequently {\em even with unbroken SUSY},
R-symmetry can be violated by higher-dimensional operators, although
only at certain orders of perturbation theory. The R-charge deficit
$\Delta r$ is related to the Euler characteristic of the string
worldsheet, $\chi=2-2g-h$, where $g$ is the genus and $h$ is the
number of boundaries: $|\Delta r|\leq - 2\chi$, with $\chi\leq 0$.
Of course, only SUSY-preserving interactions are allowed in string
perturbation theory. Note that if SUSY is broken at the string
level, these additional restrictions are lifted. As an example,
consider the disk worldsheet with $g=0,~h=1$, hence $\chi=1$, which
incorporates the effects of all tree-level interactions, including
the exchanges of virtual Regge excitations. Recall that gaugino and
gauge boson vertex operators are inserted at the disk boundary
``attached'' to the associated stack of D-branes. In this case,
$\Delta r=0$, so that two like-helicity gauginos cannot annihilate
into gauge bosons and the amplitude for $\lambda^{\pm}\lambda^{\pm}$
annihilation into an arbitrary number of gauge bosons vanishes at the disk
level, to all orders in $\alpha'=1/M_s^2$. This can be confirmed by
using SUSY Ward identities along the lines of
Ref.~\cite{Stieberger:2007jv}. {}For a gaugino pair to annihilate
into gauge bosons one needs a worldsheet with Euler characteristic
$\chi=-1$. It can be realized in two ways: a ``genus 3/2''
worldsheet with $g=1,
~h=1$~\cite{Antoniadis:2004qn,Antoniadis:2005xa}, which is
essentially a disk with a closed string handle depicted in the left-hand side of 
Fig.~\ref{fig1}, and $g=0,~ h=3$,
which is a two-loop open string
worldsheet~\cite{Antoniadis:2005sd} depicted in the right-hand side of Fig.~\ref{fig1}.
The case of a worldsheet with three boundaries is particularly
interesting \cite{Antoniadis:2005sd}.  If one inserts two vertex
operators creating gauge bosons or gauginos associated with one stack of
D-branes, say $c$, at a single worldsheet boundary, two vertex
operators associated with stack $c'$ at another boundary, while keeping
``empty'' the third boundary as in the right-hand side of Fig.~\ref{fig1}, one obtains a
non-vanishing contribution to the four-point scattering amplitude. The
corresponding effective interaction is described by the supersymmetric
F-term,
\begin{equation}
\label{lint}
{\cal L}_{\rm int}= 3 \ g_s^3  \ N\ M_s^{-3}\,\wf\,({\rm Tr}\,
W^c_{\alpha}\epsilon^{\alpha\beta}W^c_{\beta})({\rm Tr}\,
W^{c'}_{\gamma}\epsilon^{\gamma\delta}W^{c'}_{\delta})\big|_{\theta^2}
~+~ c.c.,
\end{equation}
where $W$ are the usual chiral superfields with field strengths
associated with appropriate gauge groups and the traces are taken in the
fundamental representations. Here, $g_s$ is the string coupling, and
$N$ is the number of D-branes attached to the empty boundary.  (A
total of six possibilities in the three-stack model under
consideration.) The factor of 3 is the number of choices of the empty boundary.
If the stack $c' \neq c$, the above contribution
yields the full amplitude.  The factor $F^{(0,3)} = 3N\wf$
is the genus zero
topological partition function on a worldsheet with $h=3$
boundaries. It depends on the moduli of the compact space and takes into
account various string configurations in six internal dimensions.  The
corresponding amplitudes are called ``topological'' because they are
determined by the topology of the compact dimensions and, unlike standard
amplitudes, they do {\em not} contain any kinematical singularities
associated with Regge excitations. On the other hand, if $c'=c$, the
four-point amplitude also receives non-topological contributions
from all four vertices inserted at the same
boundary~\cite{Antoniadis:1993ze}.\footnote{The amplitudes induced
by the ``genus 3/2'' worldsheets, which involve closed strings
propagating in the handle, can be analyzed in a similar way.
These amplitudes are related to (and, in some sense, they are
``square roots'' of) the genus 2 topological amplitudes
in type II string theory~\cite{Antoniadis:1993ze}. Although they
are not strictly topological, one does not expect kinematical
singularities to appear in the intermediate channels. The only
difference from the two-loop open string worldsheets is that in
the genus 3/2 case gauginos and gauge bosons must belong to the
same stack of D-branes.} In this Letter, we are mainly interested
in the former case, with all bino-like gauginos associated with the
$U(1)$ stack and gauge bosons associated with all three stacks. For
our purposes, it is sufficient to focus on the effective interaction term of
Eq.~(\ref{lint}). In the case where the annihilation occurs through emission of
a pair of $U(1)$ gauge bosons, the interaction will include  unknown
non-topological and ``genus 3/2'' contributions. Note that such a
term can be induced not only by string physics,
but also by any R-symmetry violating extension of the standard model
at a sufficiently low energy scale.

The interaction term contained in Eq.~(\ref{lint}), relevant to
$s$-wave gaugino annihilation, is
\be
\label{llambda} {\cal
  L}_{\lambda\lambda}=\frac{\cal
  T}{8}\,M_s^{-3}\,(\delta_{c_1c_2}\lambda^{c_1}_{\alpha_1}
\epsilon^{\alpha_1\alpha_2}\lambda^{c_2}_{\alpha_2})(\delta_{c'_3c'_4}
F^{c'_3\beta_3}_{\alpha_3}
\epsilon^{\alpha_3\alpha_4}\epsilon_{\beta_3\beta_4}F^{c'_4\beta_4}_{\alpha_4})
~+~ c.c.,
\ee
with lower case Latin and Greek letters labeling gauge
and spinor indices, respectively, and the dimensionless coupling
constant
\be
\label{tau}
{\cal T}=3N\,g_s^3\  \wf \, .
\ee
In Eq.~(\ref{llambda}),
$F_{\alpha}^{\beta}$ denote self-dual gauge field strengths in the
spinorial representation,
\be
\label{fspin}
F^{\beta}_{\alpha}
=\frac{1}{2}F_{\mu\nu}\sigma_{\alpha\dot\alpha}^{\mu}\bar\sigma^{\nu\dot\alpha\beta}=
F_{\mu\nu}\sigma_{\alpha}^{\mu\nu\beta}: \qquad
F_{\alpha}^{\beta}F^{\alpha}_{\beta}=-F^{\mu\nu}F_{\mu\nu}+\frac{i}{2}\,
\epsilon_{\mu\nu\rho\lambda}F^{\mu\nu}F^{\rho\lambda}.
\ee
In order
to compute the amplitude for the gaugino pair annihilation into two
gauge bosons, we need the wavefunctions of all particles in addition to the interaction term
(\ref{llambda}). We consider the
case of massive gauginos, with Majorana mass $m$, without
 addressing details of the SUSY breaking mechanism. In the center
of mass frame, two gauginos moving along the
$z$-axis with three-momenta $\bf k$ and $-\bf k$, respectively,
are described by the Majorana spinors:
\be
\label{majorana}
\bigg( {u\atop \bar u}\bigg)\quad {\rm with}\quad
u^+(\pm{\bf k})=\sqrt{m}\left( e^{\mp\eta/2}\atop 0\right)~,\quad
u^-(\pm{\bf k})=\sqrt{m}\left( 0\atop e^{\pm\eta/2}\right),
\ee
where
$u^{\pm}$ refer to spin up and down, respectively, and the rapidity
\be
\label{rapid}
\eta={\rm sinh}^{-1}\left(\frac{|\bf k|}{m}\right).
\ee
On the other hand, the polarization vectors for the gauge bosons
are \be\label{pol} \epsilon_{\mu}^{\pm}(k,q)=\pm\frac{\langle
  q^{\mp}|\gamma_{\mu}|k^{\mp}\rangle}{\sqrt{2}\langle
  q^{\mp}|k^{\pm}\rangle}\,,
\ee
where $\epsilon^{\pm}$ refer to
helicities, $k$ is the momentum and $q$ is an arbitrary reference
vector. In Eq.~(\ref{pol}) and below, we use the notation of
Ref.~\cite{Dixon:1996wi}.  Since,
\be
\label{sel}
F^{\beta}_{\alpha}|_{F_{\mu\nu}=\epsilon_{\mu}^+k_{\nu}-
  \epsilon_{\nu}^+k_{\mu}}=0\,,
\ee
the interaction term written
explicitly in Eq.~(\ref{llambda}) couples only to $(--)$ gauge boson
helicity configurations while its complex conjugate couples only to $(++)$ configurations. In
all, there are only four non-vanishing helicity amplitudes:
\be
{\cal
  M}(\lambda^{c_1}_-\lambda^{c_2}_-~\to~ g^{c'_3}_+g^{c'_4}_+)~\equiv~
{\cal M}_{--}\quad,\quad {\cal M} (\lambda^{c_1}_+\lambda^{c_2}_+~\to~
g^{c'_3}_+g^{c'_4}_+)~\equiv~ {\cal M}_{+-}\ ,\ee \be {\cal
  M}(\lambda^{c_1}_+\lambda^{c_2}_+~\to~ g^{c'_3}_-g^{c'_4}_-)~\equiv~
{\cal M}_{++}\quad,\quad {\cal M}(\lambda^{c_1}_-\lambda^{c_2}_-~\to~
g^{c'_3}_-g^{c'_4}_-)~\equiv~ {\cal M}_{-+}\ ,
\ee
where the notation
corresponds to all particles incoming.

The  effective interaction F-term (\ref{lint}) yields
\begin{equation}
{\cal M}_{--} =2\,{\cal T}M_s^{-3}\,\delta^{c_1c_2}\delta^{c'_3c'_4}\epsilon^{\alpha_1\alpha_2}
u^-_{\alpha_1}({\bf k}_1={\bf k})\, u^+_{\alpha_2}({\bf k}_2=-{\bf k})\, \epsilon_{3\mu}^- k_{3\nu}\epsilon_{4\rho}^- k_{4\lambda}
{\rm Tr}(\sigma^{\mu\nu}\sigma^{\rho\lambda})\ .
\end{equation}
Using Eqs.~(\ref{majorana}) and (\ref{pol}), we obtain
\be\label{mplus}
{\cal M}_{--} ~={\cal T}M_s^{-3}m\, e^{\eta}\langle
34\rangle^2\, ,
\ee
where we omitted the trivial $\delta^{c_1c_2}\delta^{c'_3c'_4}$ group
factor enforcing identical gauge charges of the annihilating gauginos
as well as those of the created gauge bosons.  Similarly, for the
process with fermion helicities reversed,
\begin{equation}
\label{mminus} {\cal M}_{+-} ~=-{\cal T}M_s^{-3}m\,
e^{-\eta}\,\langle 34\rangle^2\, .
\end{equation}
The two remaining amplitudes, ${\cal M}_{++}$ and ${\cal M}_{-+}$ are
obtained by complex conjugating ${\cal M}_{--}$ and ${\cal
  M}_{+-}$, respectively.

At this point we focus on one specific assignment of stacks to
boundaries. With a choice of binos (hypercharge gauge bosons) as our
LSP, and with the assumption of relatively small mixing with the other
$U(1)$ subgroups in stacks $a$ and $b$, the bino is largely associated
with the $U(1)$ stack $c$.  Under the preceding assumption of small
mixing, we note that each photon ($Z$) vertex introduces a factor of
approximately $\sin\theta_W$ $(\cos \theta_W)$ if inserted at the
boundary associated with the $U(2)$ stack $b$, and $\cos\theta_W$
$(\sin \theta_W$) if inserted at the boundary associated with stack
$c$. To retain the purely topological structure of the amplitude we
attach the second boundary to stacks $a,\ b,\ c$ and leave the third boundary
empty.

In order to compute the annihilation rate, we need the sum of squared
amplitudes, averaged over the helicities and gauge indices of initial
gauginos and summed over the helicities and gauge indices of final
gauge bosons:
\begin{eqnarray} |{\cal M} (\chi^0 \chi^0 \to WW)|^2 & = & \frac{N_{c'}^2-1}{4N_{c}^2}(\,|{\cal
  M}_{--}|^2+|{\cal M}_{+-}|^2+|{\cal
  M}_{++}|^2+|{\cal M}_{-+}|^2) \nonumber \\
&= & \frac{3}{2}\, {\cal T}^2 \,
\frac{s^2(s-2m_{\chi^0}^2)}{M_s^{6}} \ ,
\label{intoW}
\end{eqnarray}
where the Mandelstam variable, $s=(k_1+k_2)^2$, and $WW$ denotes
final states including $W^+W^-,\ ZZ,\ \gamma Z,$ or $\gamma \gamma.$
Near threshold $(s\simeq 4m_{\chi^0}^2)$, the total
annihilation rate into the three $SU(2)$ gauge vector bosons is
\begin{equation}
\left. \sigma v \right|_{WW} =  \frac{3c}{4\pi}\, {\cal T}^2
\left(\frac{\hbar}{M_s\ c}\right)^2  \rho^4  \,,
\label{sigmavWW}
\end{equation}
where $\rho\equiv m_{\chi^0}/M_s$. In a similar manner,
\begin{equation}
\left. \sigma v \right|_{gg} =  \frac{8c}{4\pi}\, {\cal T}^2
\left(\frac{\hbar}{M_s\ c}\right)^2  \rho^4  \, ,
\label{sigmavgg}
\end{equation}
and
\begin{equation}
\left. \sigma v \right|_{BB} = \zeta^2 \frac{c}{4\pi}\, {\cal T}^2
\left(\frac{\hbar}{M_s\ c}\right)^2  \rho^4  \, .
\label{sigmavBB}
\end{equation}
The factor $\zeta$ (which in principle can take any real value) parameterizes the uncertainty in the $\chi\chi\rightarrow BB$
amplitude because of
the aforementioned non-topological components in the matrix element, where all
four vertices are attached to the same boundary, or due to scattering in the ``genus 3/2''
configuration. Dominance of the topological component corresponds to $\zeta\simeq +1.$

We now constrain a combination of the free parameters of the model by requiring that neutralinos
have the measured dark matter abundance~\cite{Komatsu:2008hk}.
The density of neutralinos that survives after freezing out of thermal
equlibrium in the early universe is given by
\begin{equation}
\Omega_{\chi^0}h^2 \simeq 0.1 \bigg(\frac{x_{\rm FO}}{20}\bigg)
\bigg(\frac{g_{\star}}{80}\bigg)^{-1/2} \bigg(\frac{\langle \sigma
v \rangle_{\rm eff}}{3 \times 10^{-26} \, {\rm cm}^3/{\rm s}}\bigg)^{-1},
\end{equation}
where $x_{\rm FO}$ is the neutralino mass divided by the freeze-out
temperature, $g_\star$ is the number of external degrees of freedom
available at the freeze-out temperature, and $\langle \sigma v
\rangle_{\rm eff}$ is the effective neutralino annihilation cross
section evaluated at the freeze-out temperature. The desired effective
annihilation rate, $\left. \sigma v \right|_{WW} + \left. \sigma v
\right|_{gg} +\left. \sigma v \right|_{BB} = \langle \sigma
v\rangle_{\rm eff} \simeq 3 \times 10^{-26}$ cm$^3$/s,\footnote{In
  addition to neutralino self-annihilation, $\langle \sigma
  v\rangle_{\rm eff}$ can potentially include the effects of
  coannihilation between neutralinos and other superparticles of
  similar mass. We neglect contributions from these processes.}
required to generate the measured relic density, $\Omega_{\rm CDM} h^2
= 0.113 \pm 0.003$, is obtained if
\begin{equation}
\left(1 + 0.083(\zeta^2-1)\right)\;\left(\frac{\wf}{2.8}\right)^2 \; \left(\frac{g_s}{0.2}\right)^6\;
  \left(\frac{\rho}{0.5}\right)^4 \left(\frac{2~{\rm TeV}}{M_s}\right)^2   \simeq 1 \,.
\label{tksgivin}
\end{equation}
 A sizable value for $\wf$ is not implausible. As an example, consider the
magnetized brane model whose partition function is given by Eqs.~(5.28)-(5.30) of
Ref.~\cite{Antoniadis:2005sd}. Crudely replacing the sum over discrete lattice
momenta with integrals, one finds that $\wf$ is proportional a product of three
wrapping numbers of a D9 brane around three 2-tori. This number can in principle
be large, thus widening the available $m_{\chi^0}$--$M_s$ parameter 
space.\footnote{ This includes the range of string scales consistent with the
correct weak mixing angle found in the $U(3) \times U(2) \times U(1)$ quiver
model~\cite{Antoniadis:2000ena}. } For example, if $\wf=6$, then for
$m_{\chi^0}= 2~{\rm TeV},$ $M_s$ can be probed to 4~TeV.

These results have important implications for ongoing and future gamma ray
searches for dark matter. In particular, neutralinos annihilating in the
Milky Way halo to final states containing a photon
(such as $\gamma \gamma$ or $\gamma Z$) lead to a very distinctive
gamma-ray line which if sufficiently bright could provide a ``smoking gun''
signature of annihilating dark matter.

If we assume little mixing with $U(1)$'s from stacks $a$ and $b$,
the projection onto any photon ($Z$) in the $W^3 W^3$ final state
entails a mixing angle $\sin \theta_W$ ($\cos \theta_W$).\footnote{We note 
in passing that in the minimal extension of the
standard model this mixing angle is fixed and introduces a
multiplicative factor of 0.96 into the right-hand-side of
Eqs.~(\ref{intoW})--(\ref{sigmavBB})~\cite{Berenstein:2006pk}.} For annihilation into
the various channels we find,
\begin{eqnarray}
\left. \sigma v \right|_{\gamma \gamma} &=& \tfrac{1}{3}\WW2\left(\s2\ +\ \zeta \coseno\right)^2\,,\\
\left. \sigma v \right|_{ZZ}& = & \tfrac{1}{3}\WW2 \left(\coseno\ +\ \zeta \s2 \right)^2\,,\\
\left. \sigma v \right|_{\gamma Z}& = & \tfrac{1}{3} \WW2\ 2\coseno\s2\  \left(1 - \zeta\right)^2\,,\\
\left. \sigma v \right|_{W^+W^-} &=& \tfrac{2}{3}\WW2\,.
\label{branchings}
\end{eqnarray}
($\left. \sigma v\right|{gg}$ is given in Eq.~\ref{sigmavgg} above.)  The vanishing of $\left. \sigma v \right|_{\gamma Z}$ for $\zeta= +1$ is a reflection of the symmetry in Eq.~(\ref{lint}) in the topological case, where the coupling is independent of the choice of assignment of the stacks on the boundaries.  For $\zeta=+1$, these cross sections yield an 8.3\% 
branching fraction to $\gamma\gamma$. The $\gamma\gamma$ fraction is much larger than is predicted by the existing one-loop broken SUSY calculations~\cite{Bergstrom:1997fh,Ullio:1997ke}. For all parameter space satisfying the measured dark matter abundance~\cite{Komatsu:2008hk}, the standard annihilation rates to $\gamma \gamma$ or $\gamma Z$ are typically smaller than about $10^{-28}\ {\rm cm}^3/{\rm s}$. In contrast, our model predicts $\left. \sigma v\right|_{\gamma \gamma} \sim 3 \times 10^{-27}$ cm$^3$/s, which is more than an order of magnitude larger than the standard SUSY result.

\begin{figure}[t]
\begin{center}
\includegraphics[height=7.0cm]{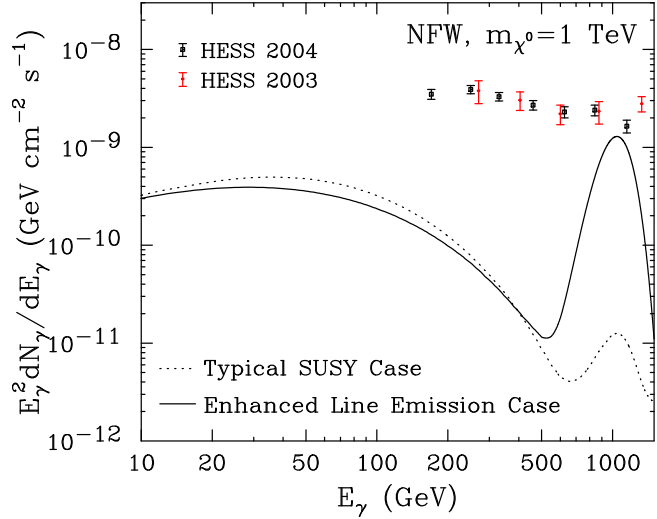}
\caption{The gamma-ray spectrum from neutralino dark matter
  annihilating in the Galactic Center (within a solid angle of
  $10^{-3}$~sr), assuming the NFW halo profile. The
  spectrum has been convolved with a gaussian of width $\Delta
  E_\gamma/E_\gamma$ =15\%, the typical energy resolution of
  H.E.S.S. and other ground based gamma-ray telescopes. The solid curve
  corresponds to dark matter annihilation with $\zeta=+1$, for which
  the $\gamma \gamma$ final state has a branching fraction of 8.3\%. The dotted curve corresponds
  to 0.1\% branching fractions to $\gamma \gamma$ and $\gamma Z$, typical of neutralino
  annihilation in the MSSM. In both cases, we considered a 1 TeV mass
  and a total annihilation cross section of $3\times 10^{-26}$
  cm$^3$/s. The continuum portion of the spectrum arises from the decay
  products of the W and Z bosons, and gluons as calculated using Pythia. Also
  shown for comparison are the 
  H.E.S.S. data~\cite{Aharonian:2004wa} which are generally interpreted to
  be of astrophysical origin~\cite{Aharonian:2004jr}.}
\label{fig3}
\end{center}
\end{figure}

For neutralinos with masses above a few hundred GeV, the H.E.S.S.
observations of the Galactic Center (GC)~\cite{Aharonian:2004wa} can be
used to probe the dark matter annihilation cross
section. The flux of gamma rays from dark matter annihilation in the
GC is given by
\begin{equation}
  \Phi_{\gamma}(E_{\gamma},\psi) \simeq \frac{\sigma v|_{i}}{8 \pi}
  \left. \frac{dN_{\gamma}}{dE_{\gamma}} \right|_i \int_{\rm l.o.s.} n_{\chi^0}^2(r)  \  dl(\psi) \ d\psi\,,
\end{equation}
where $i$ denotes the final state, $\psi$ is the angle observed away
from the GC, $dN_{\gamma}/dE_{\gamma}$ is the spectrum of gamma rays
produced per annihilation, and $n_{\chi^0}(r)$ is the number density
of dark matter particles as a function of the distance from the GC.
The integral is performed over the observed line-of-sight assuming a
dark matter distribution which follows the Navarro-Frenk-White (NFW)
halo profile~\cite{Navarro:1996gj}.  In Fig.~\ref{fig3}, the dotted
curve is the gamma-ray spectrum corresponding to a 1~TeV neutralino
with a total annihilation rate $\sigma v|_{\rm tot} = 3 \times
10^{-26}$ cm$^3$/s, that annihilates to $\gamma \gamma$ and $\gamma Z$
with branching fractions of 0.1\%, which is typical for a TeV
neutralino in the the minimal supersymmetric standard model (MSSM).
For significantly larger branching fractions to $\gamma\gamma$ or
$\gamma Z$, the prospects for detection are greatly improved. The
solid curve in Fig.~\ref{fig3}, corresponding to $\zeta = +1$, is the
gamma-ray spectrum for a neutralino that annihilates to $\gamma\gamma$
with an 8.3\% branching fraction, and does not annihilate to $\gamma
Z$. Unlike the case of a typical neutralino, a very bright and
potentially observable gamma-ray feature is predicted. For example,
the suggestive structure at 2.5~TeV in H.E.S.S.  data from
2004~\cite{Aharonian:2004wa} can be easily accommodated within this
model.  If an experiment were to detect a strong gamma-ray line
without a corresponding continuum signal from the cascades of other
annihilation products, it could indicate the presence of a low string
scale.

In summary, within the context of D-brane TeV-scale string
compactifications, we constructed a model that generates a supersymmetric
R-symmetry violating effective Lagrangian which allows for the $s$-wave
annihilation of neutralinos, once gauginos acquire mass through an
unspecified mechanism. The model allows for a neutralino relic abundance
consistent with the measured dark matter density.
The branching fractions to monochromatic gamma rays is orders
of magnitude larger than in the MSSM. 
A very bright and distinctive gamma-ray line 
that may lie within the reach of current or next-generation gamma-ray
telescopes is predicted. A flux near the limit presently imposed by the H.E.S.S. data
would strongly support a near purely topological origin for the R-symmetry
violating effective Lagrangian.\\

We thank Ignatios Antoniadis and Dan Feldman for useful
discussions. H.G. and D.M. thank the Aspen Center for Physics for
hospitality.  This research was supported by the DoE, NASA, and NSF.


\end{document}